\documentclass[runningheads,a4paper]{llncs}

\usepackage{amssymb}
\usepackage{graphicx}
\usepackage[T1]{fontenc}

\usepackage{xspace}
\newcommand{\scxp}[1]{\protect{\textsc{#1}}\xspace}
\newcommand{\coron}{\scxp{Coron}}

\usepackage{url}

\newcommand{\keywords}[1]{\par\addvspace\baselineskip
\noindent\keywordname\enspace\ignorespaces#1}

\begin{document}

\mainmatter  

\title{The \coron System}

\titlerunning{The \coron System}

%
%
\author{Mehdi Kaytoue\inst{1} \and Florent Marcuola\inst{1} 
\and Amedeo Napoli\inst{1} \and Laszlo Szathmary\inst{2} \and Jean Villerd\inst{1}
}
\authorrunning{M. Kaytoue, F. Marcuola, A. Napoli, L. Szathmary and J. Villerd}

\institute{Laboratoire Lorrain de Recherche en Informatique et ses Applications (LORIA)\\
  Campus Scientifique -- BP 239 -- 54506 Vand{\oe}uvre-l\`{e}s-Nancy Cedex (France)\\
\email{\{kaytouem, marcuolf, napoli, villerd\}@loria.fr}
\and
D\'{e}partement d'Informatique -- Universit\'{e} du Qu\'{e}bec \`{a} Montr\'{e}al  (UQAM) \\
  C.P. 8888 -- Succ. Centre-Ville, Montr{\'e}al H3C 3P8 (Canada) \\
\email{Szathmary.L@gmail.com}
}
%
%
\maketitle

\begin{abstract}
\coron is a domain and platform independent, multi-purposed data mining toolkit, 
which incorporates not only a rich collection of data mining algorithms, but 
also allows a number of auxiliary operations. To the best of our knowledge, a 
data mining toolkit designed specifically for itemset extraction and association 
rule generation like \coron does not exist elsewhere. \coron also provides support 
for preparing and filtering data, and for interpreting the extracted units of knowledge.

\end{abstract}

\section{System Overview}
Born for a particular need in a cohort study~\cite{EGC06}, \coron is now a  framework 
of knowledge discovery in databases on its own, used in several application domains, 
e.g.~\cite{Daquin07,Kaytoue08,Ignatov08}. Intended to an educational and scientific 
usage, the \coron system is articulated into several modules for preparing and mining 
binary data, and filtering and interpreting the extracted units. Thus, from binary data 
(possibly obtained from a discretization procedure), \coron allows one to extract itemsets 
(frequent, closed, generators, etc.) and then to generate association 
rules (non-redundant, informative, etc.). Building concept lattices is also possible. 
The system includes many classical algorithms of the literature, but also others that 
are specific to \coron~\cite{szathmary07c,DS08,IDA09}. The software is freely 
available at \url{http://coron.loria.fr}. Mainly  written in Java, \coron is compatible 
with  the Unix, Mac and Windows operating systems and is of command-line usage.

\section{A Global Data Mining Methodology}
The methodology was initially designed for mining biological cohorts, but it is 
generalizable to any kind of database. It is important to notice that the whole process 
is guided by an expert, who is a specialist of the domain related to the database. His 
role may be crucial, especially for selecting the data and for interpreting the extracted 
units, in order to fully turn them into knowledge units. In our case, the extracted knowledge 
units are mainly association rules. At the present time, finding association rules is one of 
the most important tasks in data mining. Association rules allow one to reveal ``hidden'' 
relationships in a dataset. Finding association rules requires first the extraction of 
frequent itemsets.

\begin{figure}[b!] \centering
\includegraphics[width=1.0\columnwidth]{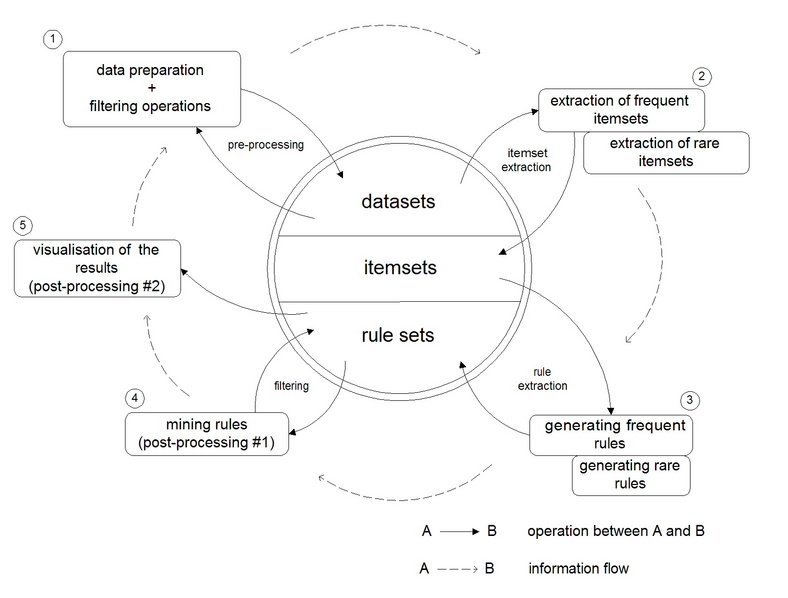} 
\caption{Architecture of the \coron System}
\label{loop}
\end{figure}

The methodology consists of the following steps: Definition of the study framework; Iterative step: data preparation and cleaning, pre-processing step, processing step, 
post-processing step; Validation of the results and Generation of new research hypotheses; 
Feedback on the experiment. The life-cycle of the methodology is shown in Figure \ref{loop}. 
Coron is designed to satisfy the present methodology and offers all the tools that are necessary 
for its application in a single platform. 

\paragraph{Pre-processing.}
These modules propose several tools for manipulating and formatting large data.  The data are 
described by binary tables in a simple text-file format: some individuals in lines possess or 
not some properties in column. The main possible operations are: (i) discretization of numerical 
data, (ii) conversion of different file formats, (iii) creation of the complement of the binary 
table, and (iv) other projection operations such as transposition of the table.

\paragraph{Data mining.}
Extracting itemsets and association rules is a very popular task in data mining. Concept 
lattices are mathematical structures supported by a rich and well established formalism,
namely, Formal Concept Analysis~\cite{ganter99}. A concept lattice is represented by a 
diagram giving nice visualization of classes of objects of a domain. Thus, the data mining modules of the \coron System offer the following possibilities:

\begin{itemize}
\item Itemset extraction: frequent, closed, rare, generators, etc. 
This task is performed by a large collection of algorithms based on different 
search strategies (depth-first, level-wise, etc.).
\item Association rules generation: frequent, rare, closed, informative, minimal
non-redundant, Duquenne-Guigues basis, etc. 
These rules are given with a set of measures such as support, confidence, lift, conviction, etc.
\item Concept lattice construction.
\end{itemize}

\paragraph{Post-processing.}
Extracted units from the data mining step may be very numerous, and hide some units of
higher interest. Thus, \coron proposes some filtering operations that should be done
in interaction with a domain expert. The analyst may filter rules w.r.t. the length of
its components, and/or the presence of a given property. He may also retain the $k$ best
extracted units w.r.t. a measure of interest. It is also possible to color some properties
of a list of association rules.

\paragraph{Toolbox.} Finally, auxiliary modules allow one to visualize equivalence classes
of itemsets, randomly generate binary data, etc.
\section{Applications}
\coron has been used for the following tasks: extraction of knowledge of adaptation in
case-based reasoning~\cite{Daquin07}, gene expression data analysis~\cite{Kaytoue08,Kaytoue09},
information retrieval~\cite{Coria08}, recommendations for internet advertisement~\cite{Ignatov08},
biological data integration~\cite{Coulet08}, and finally, cohort studies~\cite{EGC06}.
\section{Work in Progress}
Currently, we are studying how to integrate \coron in platforms using graphical data-flows,
such as Knime \cite{knime}, whose popularity is increasing (\url{http://www.knime.org}). This
would allow \coron to interact with many other useful tools, most importantly avoiding a
command-line usage. Also, other tools will be integrated in \coron to consider complex data,
mainly numerical, see e.g.~\cite{Kaytoue09}. Finally, we have recently set up a forum to gather
questions, comments and suggestions from \coron users (\url{http://coron.loria.fr/forum/}). 

In this paper, we have given a brief overview of the \coron System. For more details, please
refer to the project's website at \url{http://coron.loria.fr}.
\section*{Acknowledgements} 
The authors would like to thank the following persons for their participation in the
development of \coron: F. Collignon, B. Ducatel, S. Maumus, P. Petronin, T. Bouton, A. Knobloch, N. Sonntag, Y. Toussaint.

\end{document}